\documentclass[fleqn,usenatbib]{mnras}
\usepackage{newtxtext,newtxmath}
\usepackage{pdflscape} 
\usepackage{afterpage}

\usepackage[T1]{fontenc}
\usepackage{ae,aecompl}

\usepackage{graphicx}	
\usepackage{amsmath}	
\usepackage[usenames,dvipsnames]{xcolor}
\usepackage[export]{adjustbox} 

\defcitealias{FFtech}{Paper~I}
\defcitealias{Hopwood15}{Hop15} 
\defcitealias{FFredshift}{Paper~II}
\defcitealias{FFlineID}{Paper~III}
\defcitealias{FFncc}{Paper~IV}

\title[SPIRE FF V: Rotation of NGC~891]{
The \textit{Herschel} SPIRE Fourier Transform Spectrometer Spectral Feature Finder V. Rotational measurements of NGC~891 \thanks{\textit{Herschel} was an ESA space observatory with science instruments provided by European-led Principal Investigator consortia and with important participation from the National Aeronautics and Space Administration.}}
\author[C. S. Benson et al.]{
Chris S. Benson,$^{1}$\thanks{E-mail: chris.benson@uleth.ca}
L.~D. Spencer,$^{1}$
I. Valtchanov,$^{2}$
J. Scott,$^{1}$
N. H\l{}adczuk$^{3, 4}$
\\
$^{1}$Institute for Space Imaging Science, Department~of Physics \& Astronomy, University of Lethbridge, 4401 University Drive, \\
Lethbridge, Alberta, T1K 3M4, Canada \\
$^{2}$Telespazio Vega UK for ESA, European Space Astronomy Centre, Operations Department, 28691 Villanueva de la Ca\~nada, Spain\\
$^{3}$ European Space Astronomy Centre, ESA, Camino Bajo del Castillo, 28692, Villanueva de la Ca$\tilde{n}$ada, Madrid, Spain \\
$^{4}$ Gran TeCan, S.A., Instituto de Astrof\'{i}sica de Canarias, C/ V\'{i}a L\'{a}ctea, S/N, 38205 - San Crist\'{o}bal de La Laguna, S/C de Tenerife, Spain\\
}

\date{Accepted 11/2020. Received 10/2020; in original form 10/2020}

\pubyear{2020}

\begin{document}
\label{firstpage}
\pagerange{\pageref{firstpage}--\pageref{lastpage}}
\maketitle


\begin{abstract}
The ESA \textit{Herschel} Spectral and Photometric Imaging Receiver (SPIRE) Fourier Transform Spectrometer (FTS) Spectral Feature Finder (FF) project is an automated spectral feature fitting routine developed within the SPIRE instrument team to extract all prominent spectral features from all publicly available SPIRE FTS observations. In this work, we demonstrate the use of the FF information extracted from three observations of the edge-on spiral galaxy NGC~891 to measure the rotation of \ion{N}{II} and \ion{C}{I} gas at Far-infrared frequencies in complement to radio observations of the $[$\ion{H}{I}$]$ 21cm line and the CO(1-0) transition as well as optical measurements of H$\alpha$. We find that measurements of both \ion{N}{II} and \ion{C}{I} gas follow a similar velocity profile to that of \ion{H}{I} and H$\alpha$ showing a correlation between neutral and ionized regions of the interstellar medium (ISM) in the disk of NGC~891. 
\end{abstract}




\section{Introduction}
The \textit{Hershcel Space Observatory$^\ast$} is an observatory class mission of the European Space Agency (ESA) \citep{Pilbratt10} that completed four years of observations exploring the far-infrared (FIR) and submillimeter (sub-mm) Universe in April 2013 with the depletion of its liquid cryogens \citep{herschelEnd}. The Spectral and Photometric Imaging REceiver (SPIRE) was one of three focal plane instruments on board \textit{Herschel}, consisting of both an imaging photometric camera and an imaging Fourier Transform Spectrometer (FTS) \citep{Griffin10}. The SPIRE FTS has two bolometer detector arrays, the Spectrometer Long Wavelength (SLW) and the Spectrometer Short Wavelength (SSW), that simultaneously cover a frequency band of 447--1546 GHz (SLW: 447--990 GHz, SSW: 958--1546 GHz). SPIRE FTS observations provide a wealth of molecular and atomic fine-structure spectral lines including the $[$\ion{N}{II}$]$ $^3$P$_1$--$^3$P$_0$, $[$\ion{C}{I}$]$ $^3$P$_2$--$^3$P$_1$, and $[$\ion{C}{I}$]$ $^3$P$_1$--$^3$P$_0$ transitions. During \textit{Herschel's} mission, the SPIRE FTS instrument made three high resolution ($\Delta \nu \sim 1.2$ GHz) spectral observations of the spiral galaxy NGC~891 (observation IDs 1342224765, 1342224766, and 1342213376) that are publicly available through the \textit{Herschel Science Archive} (HSA)\footnote{\url{http://archives.esac.esa.int/hsa/whsa/}} \citep{spire_handbook}. The $[$\ion{N}{II}$]$ $^3$P$_1$--$^3$P$_0$ line has been measured with exceptionally high signal-to-noise ratios (SNRs) in these observations which also contain lower SNR CO and $[$\ion{C}{I}$]$ features.

Recently the SPIRE FTS observations have become more accessible through the SPIRE Spectral Feature Finder Catalogue\footnote{\url{https://www.cosmos.esa.int/web/herschel/spire-spectral-feature-catalogue}}, which includes a collection of significant spectral features extracted from all publicly available high resolution (HR) single-pointing and mapping observations by the automated SPIRE Feature Finder (FF) routine \citep{FFtech,FFredshift,FFlineID,FFncc}\footnote{Due to repeated referencing of \citet{FFtech}, \citet{FFredshift}, \citet{FFlineID}, and \citet{FFncc}, the abbreviations \citetalias{FFtech}, \citetalias{FFredshift}, \citetalias{FFlineID}, and \citetalias{FFncc}, respectively, are used in further text.}. The full SPIRE Automated Feature Extraction Catalogue (SAFECAT), contains the central frequency and SNR of $\sim165\,000$ features at SNRs greater than 5 (some lower SNR features may be found by the line identification routine, see \citetalias{FFtech, FFlineID}) from all publicly available SPIRE FTS observations in the HSA. From the spectral content of each observation and through literature cross references, radial velocity estimates for each observation are provided \citepalias{FFredshift}. Initial line identification estimates are also catalogued \citepalias{FFlineID} and the FF also employs a subroutine to better detect $[$\ion{C}{I}$]$ features that can be difficult to detect in automated line fitting routines of SPIRE spectra \citepalias{FFncc}.

To demonstrate the utility of the FF and its catalogue, we have used the central line frequencies catalogued by the FF from the three HR SPIRE FTS observations of NGC~891 to measure the rotational kinematics of ionized nitrogen and neutral carbon near the plane of the galaxy at FIR wavelengths. This data is in complement to H$\alpha$, CO(1-2), and $[$\ion{H}{I}$]$\,21\,cm measurements presented in \citet{Kamphuis2007,Garcia1992,Scoville1993,Sofue1996}, and \citet{Kijeong2011} in order to associate different energy regimes of the interstellar medium (ISM) in NGC~891. 
\vspace{-18pt}
\section{Observational Information}
NGC~891 is a nearby ($D\sim9.5$ Mpc for $H_0 = 75$ km\,$s^{-1}$\,Mpc$^{-1}$) edge-on spiral galaxy (inclination angle $\gtrsim 89^\circ$) and is similar in structure to the Milky Way \citep{vanDerKuit1981}. CO observations of the galaxy have shown a 4\,kpc molecular ring and a high velocity disk surrounding the nucleus \citep{Garcia1992, Scoville1993, Sofue1996}. $[$\ion{H}{I}$]$ measurements of this galaxy are more extend than CO measurements ($R>10$\,kpc) and both CO and $[$\ion{H}{I}$]$ measurements show some asymmetries with the south end of the galaxy being more extended than the north end \citep{Sancisi1979, Sofue1996, Kijeong2011}. H$\alpha$ measurements also find asymmetries in NGC~891 with emission being more intense in the south end of the galaxy in both its disk and halo; this suggests that star formation is more active there provided that the ionized gas is brought up into the halo by star formation processes \citep{Kamphuis2007}.

Two of the SPIRE FTS observations of NGC~891 are sparsely sampled single-pointing observations (1342224766 and 1342224765, hereafter referred to as \textbf{N}orth and \textbf{S}outh, respectively). These observations were taken on 2011-07-26 (operational day 804) and both have an integration time corresponding to 55 FTS HR scans. The FF has catalogued lines from both the extended and point source calibrations of the SPIRE FTS for these observation and the extended source calibration minimizes the gap in the SLW--SSW overlap region (see \citealt{spire_handbook,Wu2013}). The third observation of NGC~891 is an intermediately sampled mapping observation (1342213376, hereafter referred to as \textbf{M}) corresponding to four jiggle positions of the SPIRE FTS Beam Steering Mirror (BSM). Mapping observations of the SPIRE FTS are projected onto hyper-spectral cubes and the FF extracts lines from the spectra from each pixel separately (see \citetalias{FFtech}). This observation was taken on 2011-01-28 (operational day 625) with an integration time at each jiggle position corresponding to 32 FTS HR scans. The FF only extracts lines from an extended source calibration for mapping observations \citepalias{FFtech}. All three observations have been reduced by the SPIRE calibration tree \texttt{spire\_cal\_14\_3} and the emission from NGC~891 does not fill the entirety of the SPIRE beam for each detector \citep{Hughes2015}. Since emission from the galaxy is also not uniformly point-like for each detector, a proper determination of line-flux values requires semi-extended considerations outlined in \citet{Wu2013}. 

\begin{figure}
	\includegraphics[trim = 1mm 0mm 0mm 0mm, clip, width=\columnwidth]{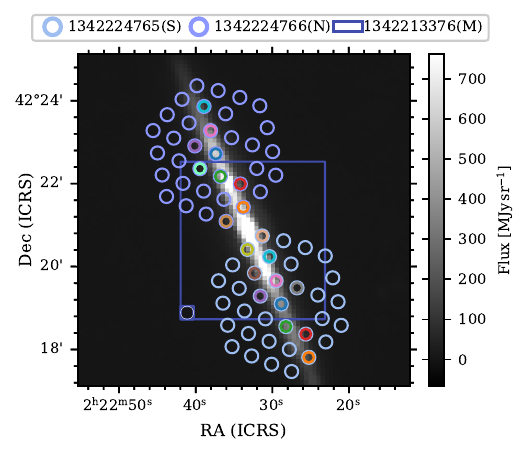}
	\vspace{-22pt}
    \caption{The SSW array of the FTS for each sparse observation (N and S) is shown with the footprint of the SSW cube from the mapping observation (M). These are imposed on a SPIRE photometer short wavelength (PSW) map of NGC~891 centered at 250$\mu$m. Each circle shows the nominal full width at half maximum (FWHM) of the SPIRE FTS beam \citep{Makiwa2013} and detectors that have observed the $[$\ion{N}{II}$]$ feature have a thicker outline with colours assigned to each unique detector. The central rectangle indicates the extent of observation M with the small inset rectangle and circle showing the on-sky size of a pixel in the associated hyperspectral cube and SPIRE SSW beam.}
	\label{fig:ObsFootPrint}
	\vspace{-12pt}
\end{figure}

\begin{figure*}
	\includegraphics[trim = 0mm 0mm 0mm 0mm, clip, width=\textwidth]{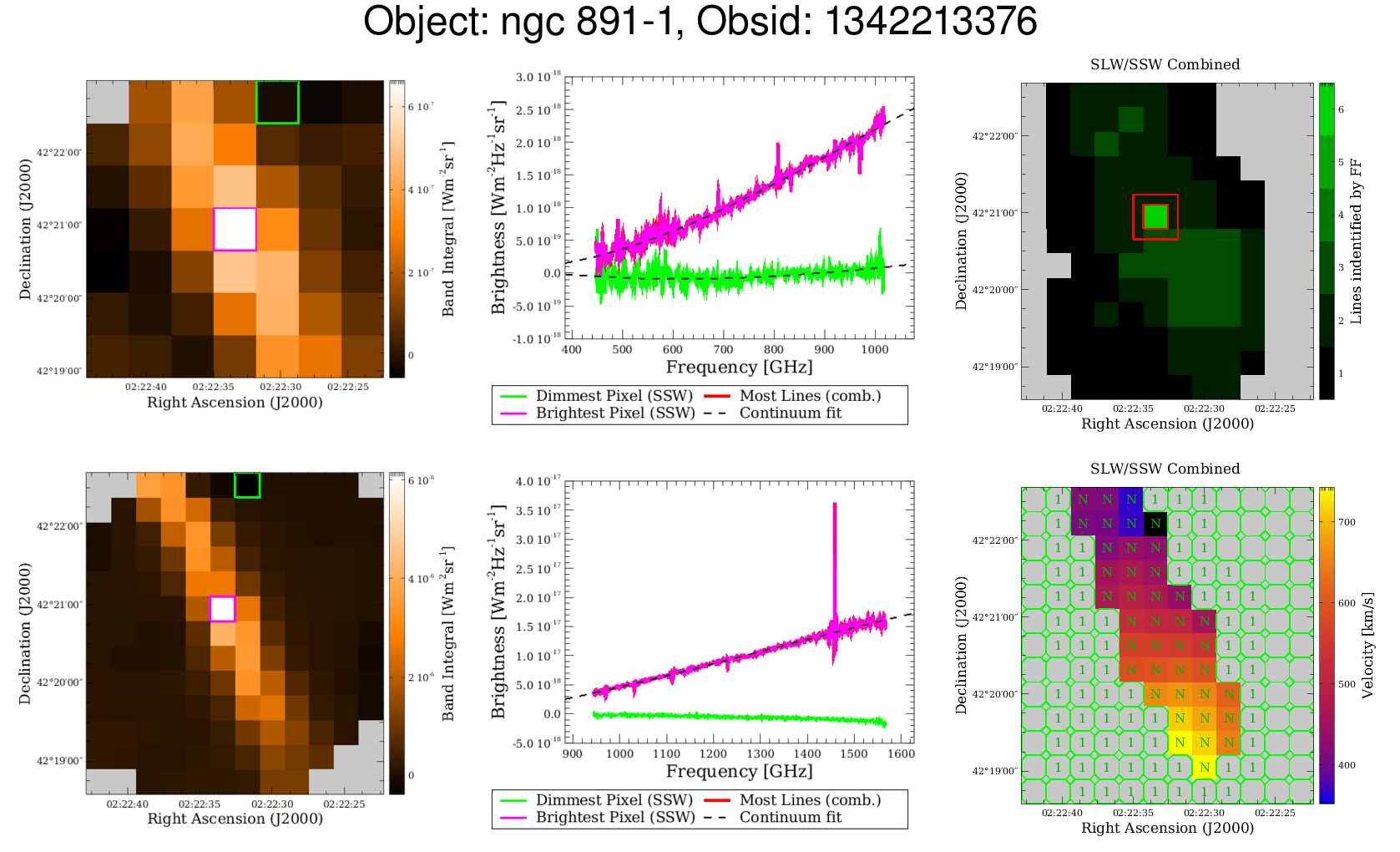}
	\vspace{-18pt}\caption{A mapping postcard of observation M as provided in the main FF catalogue \citepalias{FFtech}. These postcards are more fully described in \citetalias{FFtech}. The first column shows the intensity of each pixel integrated across each spectrometer band (SLW top, SSW bottom). The central column displays a few spectra of interest including the brightest and dimmest pixels in the cube as well as the pixel with the most lines extracted by the FF (SLW top, SSW bottom). The third column shows the number of lines discovered by the FF in each pixel (top) as well as the velocity estimate obtained for each pixel (bottom, see \citetalias{FFredshift}). In the velocity map, pixels that do not have reliable velocity estimates are coloured grey and those that have lines extracted by the FF are outlined in green. The dark green number on each pixel indicates the number of CO features found by the velocity estimate routine, an `N' character indicates when the $[$\ion{N}{II}$]$ $^3$P$_1$--$^3$P$_0$ feature is used for a velocity estimate in the FF. A full list of lines extracted from every pixel is provided in Tables \ref{tab:fullLineRes} and \ref{tab:fullLineResMap}.} 
	\label{fig:mappingPost}
	\vspace{-17pt}
\end{figure*}

The footprint of each SSW detector in observations N and S are shown with that of the SSW cube from observation M in Fig.\,\ref{fig:ObsFootPrint}. It is important to note that the full extent of observation M is not perfectly rectangular (see Fig.\,\ref{fig:mappingPost}) and that intermediately sampled mapping observations only provide 16 arcsecond ($\sim$ 1 beamwidth) sampling rather than Nyquist resolved spatial sampling of the observed region \citep{spire_handbook}. Sparse observations have a 32 arcsecond sampling determined by the detector spacing of the FTS \citep{spire_handbook}. In Fig.\,\ref{fig:ObsFootPrint} the spatial resolution of each observation is shown by the FWHM of the SSW beam for each detector in observations N and S. The pixel size of the hyperspectral cube in observation M is shown by the inset rectangle in the lower left of its footprint. Each pixel is only spatially sampled by the SPIRE beam once.

In all three observations, the prominent $[$\ion{N}{II}$]$ $^3$P$_1$--$^3$P$_0$ feature is readily detected by the SSW array. SLW detectors measure a number of neutral carbon fine-structure lines and the occasional CO rotational feature albeit at a much lower SNR. The full results from the FF line fitting are provided in Tables \ref{tab:fullLineRes} and \ref{tab:fullLineResMap} along with their molecular and atomic transitions determined by template matching in the FF's line identification routine (see \citealt{FFlineID}). The extent of $[$\ion{N}{II}$]$ features extracted by the FF improves upon the spatial coverage of $[$\ion{N}{II}$]$ emission from these observations presented in \citet{Hughes2015}. Some lines extracted by the FF remain unidentified by the FF routine, often due to highly uncertain velocity estimates for their respective detector/pixel (see \citetalias{FFredshift}) or the lines being absent from the identification template (see \citetalias{FFlineID}). A few of these have been manually identified in this work.

\begin{figure*}
\includegraphics[trim = 0mm 2mm 0mm 8mm, clip, width=\columnwidth]{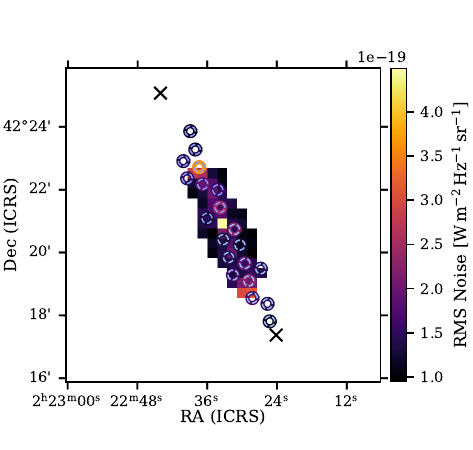}
\includegraphics[trim = 0mm 2mm 0mm 8mm, clip, width=\columnwidth]{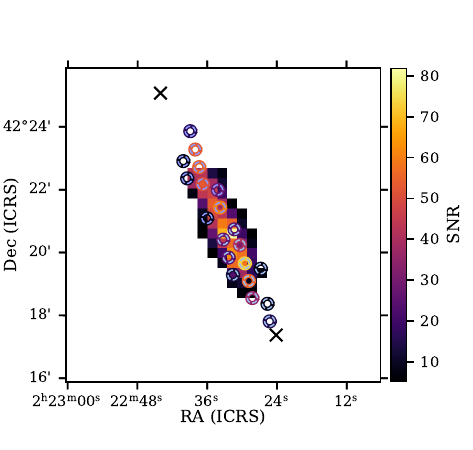}
\includegraphics[trim = 0mm 1mm 0mm 4mm, clip, width=\columnwidth]{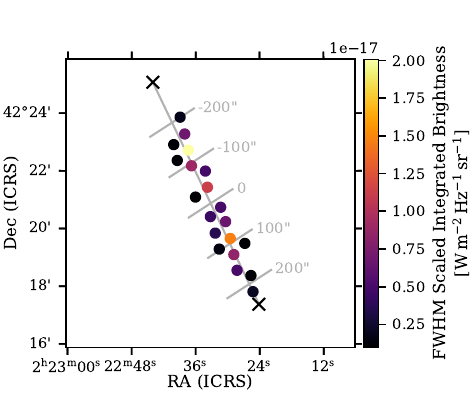}
	\vspace{-10pt}\caption{The noise map of the $[$\ion{N}{II}$]$ $^3$P$_1$--$^3$P$_0$ feature is shown for all three observations (upper left panel) with a map of the SNR (upper right panel) and a map of the integrated brightness of line scaled by its FWHM (lower panel). In each panel, the radial extent of H$\alpha$ emission measured by \citet{Kamphuis2007} is shown by the `x' markers at each extrema. $[$\ion{H}{I}$]$ emission extends past 450 arcseconds on each side (see \citealt{Kijeong2011}). For full details on noise calculations, see \citetalias{FFtech}.} 
	\label{fig:noiseMaps}
	\vspace{-17pt}
\end{figure*}

Fig.~\ref{fig:noiseMaps} demonstrates the extent of $[$\ion{N}{II}$]$ $^3$P$_1$--$^3$P$_0$ observations from all three observations compared to that of H$\alpha$ measurements made by \cite{Kamphuis2007} at much higher spatial resolution. The noise in the $[$\ion{N}{II}$]$ feature is also shown. Noise is determined as the root-mean-square (RMS) of the residual SSW spectrum over a small frequency region on either side of the $[$\ion{N}{II}$]$ feature after lines detected by the FF are removed (see \citetalias{FFtech}). The $[$\ion{N}{II}$]$ feature is detected with high SNR at distances as much as $\sim 54$ arcseconds perpendicular to the disk of the galaxy. The high error in the brightest pixel of observation M is likely due to low SNR absorption features in this spectrum that are not catalogued or extracted by the FF and remain unsubtracted from the spectrum in noise calculations (see Fig.~\ref{fig:mappingPost}). Absorption features are not catalogued by the FF unless they are of an $|$SNR$|$ greater than 10 in order to limit false detections by the automated routine (see \citetalias{FFtech}; note that absorption features are assigned a negative SNR by the FF). High SNR CO measurements within these data are limited to 6 detectors from observations N and S and a single pixel in observation M. Consequently, these sparse CO measurements are not considered further in this work. $[$\ion{C}{I}$]$ features are much more readily detected at higher SNR (see Tables \ref{tab:fullLineRes} and \ref{tab:fullLineResMap}). The integrated intensity of the $[$\ion{N}{II}$]$ feature divided by its FWHM is also shown in Fig.~\ref{fig:noiseMaps} for both sparsely sampled observations (N and S).

Neutral carbon (see \citetalias{FFncc}) and ionized nitrogen provide complementary measurements to CO and $[$\ion{H}{I}$]$ measurements of NGC~891 that probe different energy regimes within the galaxy. Neutral carbon will ionize at 11.26\,eV \citep{ionizationEnergies} but will be readily bonded into CO at energies of 11.1\,eV \citep{dysonISM, bondDis} thus $[$\ion{C}{I}$]$ features originate from a very narrow energy range separating the ionized and molecular phases of the ISM. The first ionization energy of nitrogen is 14.52\,eV \citep{ionizationEnergies} thus the $[$\ion{N}{II}$]$ $^3$P$_1$--$^3$P$_0$ feature provides a method to trace ionization regions \citep{ReviewHaffner}.
\vspace{-18pt}
\section{Results and Discussion}

\begin{figure}
	\includegraphics[trim = 0mm 0mm 0mm 0mm, clip, width=\columnwidth]{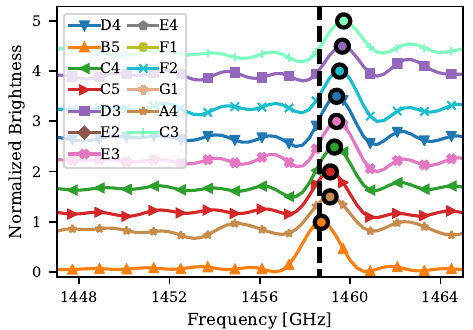}
	\includegraphics[trim = 0mm 0mm 0mm 0mm, clip, width=\columnwidth]{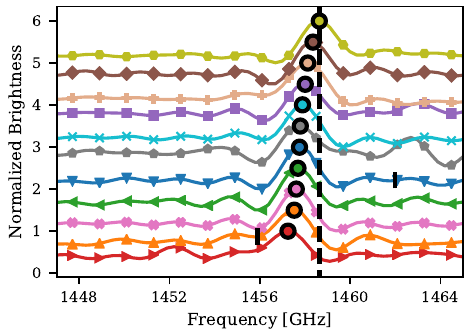}
	\vspace{-18pt}\caption{A section from the continuum subtracted spectra from each SSW detector that observed the $[$\ion{N}{II}$]$ $^3$P$_1$--$^3$P$_0$ feature in observations N (top) and S (bottom). Each detector matches the assigned colour in Fig.\,\ref{fig:ObsFootPrint} and an artificial vertical offset is applied to each spectrum. Spectra have been normalized to the peak amplitude of their respective $[$\ion{N}{II}$]$ feature. The rest-frame frequency of the $[$\ion{N}{II}$]$ feature in the local standard of rest for NGC~891 is marked by the dashed vertical line. $[$\ion{N}{II}$]$ lines extracted by the FF are marked by circles at their peak intensity while other FF lines are marked by a vertical black dash.}
	\label{fig:NIIspecs}
	\vspace{-18pt}
\end{figure}

The $[$\ion{N}{II}$]$ $^3$P$_1$--$^3$P$_0$ is detected at exceptionally high SNR in most spectra (detected at an average FF calculated SNR of 30, see \citetalias{FFtech,FFlineID}) making it an excellent tool to measure the rotation of ionized gas in the galaxy at FIR frequencies. Fig.\,\ref{fig:NIIspecs} shows the $[$\ion{N}{II}$]$ feature from each detector that observed it in observations N and S. The continuum has been extracted using the fitted continuum parameters provided by the FF. This figure also demonstrates the sinc-like line profile of the SPIRE FTS (see \citealt{spire_handbook}, \citealt{Naylor:16}). Due to this line shape, it can often prove difficult to extract low SNR lines via visual inspection which is one of the motivations for the FF project. 

\begin{figure}
	\includegraphics[trim = 0mm 0mm 0mm 0mm, clip, width=\columnwidth]{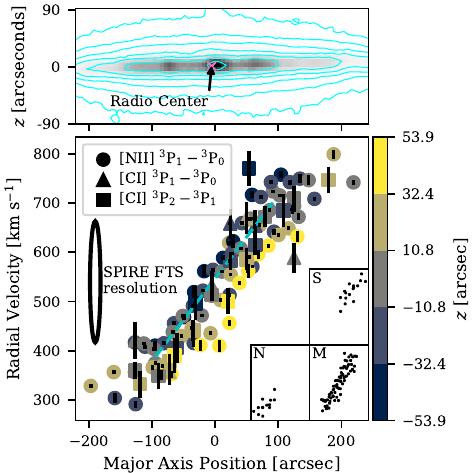}
	\vspace{-17pt}
	\caption{The top panel shows a section of the SPIRE PSW photometer map focused on NGC~891 from Fig.\,\ref{fig:ObsFootPrint} in negative color. Contour levels are 10, 21.5, 46.4, 100, 215.4, 464.2, and 1000 Mj\,sr$^{-1}$ and the radio center of the galaxy reported by \citet{Kijeong2011} is marked by the pink `x'. The bottom panel displays velocity measurements from FF catalogued $[$\ion{N}{II}$]$ and $[$\ion{C}{I}$]$ lines as a position-velocity diagram along the axis of the galaxy. Points are coloured by their orthogonal distance from the axis of the galaxy, $z$. The dashed cyan line shows the Keplerian curve fit to the center of the rotation curve in Fig.\,\ref{fig:rotCurve}. Insets in the bottom panel show the observation each set of points correspond to. Both panels share a common axis position.} 
	\label{fig:Velcurve}
	\vspace{-16pt}
\end{figure}

Fig.\,\ref{fig:Velcurve} shows the heliocentric radial velocity of the galaxy along its major axis from all three observations (N,S,M). Each point represents a radial velocity measurement from a $[$\ion{N}{II}$]$ $^3$P$_1$--$^3$P$_0$, $[$\ion{C}{I}$]$ $^3$P$_1$--$^3$P$_0$, or $[$\ion{C}{I}$]$ $^3$P$_2$--$^3$P$_1$ feature. Measurements corresponding to each observation are shown by the insets in the lower panel. The SPIRE resolution is shown in position-velocity space at the rest frequency of the $[$\ion{N}{II}$]$ $^3$P$_1$--$^3$P$_0$ feature determined by the FWHM of the SPIRE SSW beam \citep{Makiwa2013} and the 1.2\,GHz frequency resolution of the spectrometer \citep{spire_handbook}. In reality, the frequency calibration of the SPIRE FTS has been shown to allow the accurate measurement of line centers up to a factor of 1/50 the resolution of the FTS \citep{Swinyard2014, Spencer:15}; however the spectral resolution of the SPIRE FTS does not allow for the study of the internal structure of detected $[$\ion{N}{II}$]$ lines. Velocity error is determined by the error in the central line frequency determined by the line fitting of the FF. $[$\ion{C}{I}$]$ features are detected by the SLW array which has a beam that is 14--25 arcseconds greater than the SSW array. The axis of the galaxy is defined by a linear fit of all points in the SPIRE PSW map (see Fig.\,\ref{fig:ObsFootPrint}) that are at least 2.5\% of the peak intensity. This cutoff was chosen based upon visual inspection of the photometer data. Error in these positional measurements based upon this axis was estimated using a Monte-Carlo simulation varying the 2.5\% intensity cutoff 5000 times with a standard deviation of 0.3\% and again with a cutoff centered at 20\% of the peak intensity with a standard deviation of 1\%. Both simulations provided statistically equivalent results. We have defined the FIR center of the galaxy by taking the median coordinates of all pixels in the SPIRE PSW map that have $\geq 70$\% the flux of the pixel with the greatest flux (the median coordinates of pixels containing the bulge). This places the photometric center at coordinates $\alpha=2\text{h}22\text{m}33.189\text{s}$ $\delta = 42^\circ20' 52.583"$, $\sim 8$ arcseconds from the radio center reported by \citet{Kijeong2011} (see the top panel of Fig.\,\ref{fig:Velcurve}). This photometric center agrees with the brightest emission of the $[$\ion{N}{II}$]$ feature within the SPIRE spectrometer's spatial resolution for observation M. The offset between ionized nitrogen emission and atomic hydrogen is in agreement with the offset between ionized and atomic hydrogen noted by \citet{Kamphuis2007}. Though the pixel size of SPIRE PSW maps is 6 arcseconds it should be noted that the spatial beam's FWHM is $\sim 18.2$ arcseconds \citep{spire_handbook}.

The halo of NGC~891 is known to contain slower rotating \ion{H}{I} gas and diffuse ionized gas up to altitudes greater than 4\,kpc from the disk \citep{Swaters1997, Rand1990} and thus there is potential for slower rotating $[$\ion{N}{II}$]$ emission that can contaminate SPIRE observations of the disk. \citet{Swaters1997} have shown a significant decrease in the rotational speeds of \ion{H}{I} gas at $30<|z|<60$ arcseconds. With the 16.6 arcsecond FWHM SPIRE beam at the rest frequency of the $[$\ion{N}{II}$]$ $^3$P$_1$--$^3$P$_0$ line, any observations with an elevation ($|z|$) greater than $\sim 20$ arcseconds is likely subject to a significant amount of contamination from slow-moving halo gas. Based upon 16 SPIRE $[$\ion{N}{II}$]$ measurements that are within 20 arcseconds of the galaxy's disk and are less then a beam-width apart we have found that their radial velocities may disagree by 22--64\,km\,s$^{-1}$. Fig.\,\ref{fig:Velcurve} demonstrates that a single spectral resolution element of the FTS is large in velocity space; this coupled with the spatial resolution of the instrument may result in the FF reported line centers being subject to line-blending from $[$\ion{N}{II}$]$ features at multiple velocities. \citet{Keppel1991} have studied the kinematics of ionized hydrogen perpendicular to the major axis of NGC~891 and have shown that radial velocities measured by H$\alpha$ emission may decrease by as much as $\sim$55\,km\,s$^{-1}$ within perpendicular distances as wide as the SPIRE SSW beam.

Our measurements of the galactic rotation from the $[$\ion{N}{II}$]$ $^3$P$_1$--$^3$P$_0$ feature follow a similar trend to the higher spatially resolved H$\alpha$ measurements made by \cite{Kamphuis2007} and $[$\ion{H}{I}$]$ measurements presented in \citet{Sancisi1979} and \citet{Kijeong2011}, showing no major deviations from these within SPIRE's limited spatial resolution. Keplerian rotation is seen within the extent of SPIRE observations with the furthest measurements suggesting a flattening of the rotation curve due to dark matter. The limited extent of the SPIRE FTS observations ($R<11.5$\,kpc) does not allow us to determine if the $[$\ion{N}{II}$]$ emission experiences the same asymmetry as $[$\ion{H}{I}$]$ emission that occurs at $R>14$\,kpc for the south end of the galaxy \citep{Sancisi1979}. The limited spatial resolution and extent of SPIRE FTS observations do not allow us to determine whether $[$\ion{N}{II}$]$ emission follows the rigid body rotation observed in H$\alpha$ measurements that is due to the optical thickness of gas along the disk of the galaxy \citep{Kamphuis2007}, however the asymmetric decrease in signal from H$\alpha$ emission in the southern 100--200 arcseconds region of the galaxy also noted by \citet{Kamphuis2007} is apparent (see Fig.~\ref{fig:noiseMaps}).

Measurements of $[$\ion{C}{I}$]$ features near the center of the galaxy do not show the characteristic profile of the rapidly rotating molecular disk near the galactic nucleus \citep{Garcia1992, Sofue1996} and are instead more in agreement with $[$\ion{H}{I}$]$ measurements. This provides further evidence that this central disk is completely molecular. It should be noted that SPIRE measurements of $[$\ion{C}{I}$]$ features in the galaxy tend to be of much lower SNR than measurements of the $[$\ion{N}{II}$]$ $^3$P$_1$--$^3$P$_0$ feature and thus are subject to a much higher degree of uncertainty. These lines are also detected in the SLW band and their spatial resolution is significantly worse than measurements of the $[$\ion{N}{II}$]$ $^3$P$_1$--$^3$P$_0$ feature.

\begin{figure}
	\includegraphics[trim = 0mm 0mm 0mm 0mm, clip, width=\columnwidth]{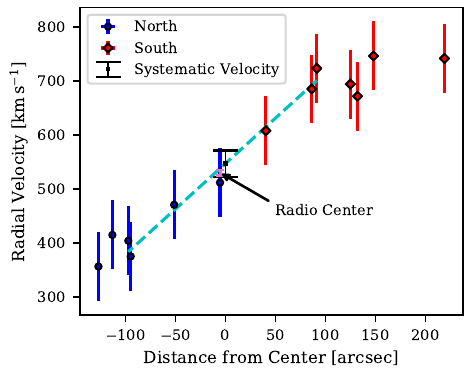}
	\vspace{-18pt}
	\caption{The rotation curve of NGC~891 measured by $[$\ion{N}{II}$]$ $^3$P$_1$--$^3$P$_0$ fine structure lines. The dashed cyan line shows a Keplerian curve fit to the central region of the rotation curve. The pink error bars show the systematic velocity at the radio center reported by \citet{Sancisi1979} and \citet{Kijeong2011} while the thin black error bars show the systematic velocity 547$\pm$24\,km\,$s^{-1}$ determined from the rotation curve measured by SPIRE detected $[$\ion{N}{II}$]$ features.}
	\label{fig:rotCurve}
	\vspace{-17pt}
\end{figure}

The rotation curve in Fig.\,\ref{fig:rotCurve} is calculated from spectra containing the $[$\ion{N}{II}$]$ $^3$P$_1$--$^3$P$_0$ feature that are within a half beam-width of the galaxy's major axis. Radial velocity has been corrected for the slight inclination of the galaxy. Within the extent of near-disk SPIRE observations, the motion is Keplerian and in agreement with the rotation curve for neutral hydrogen measured by \citet{Sancisi1979}. The large uncertainty in velocity values is indicative of the maximum velocity discrepancy (64\,km/s) between complementary measurement of $[$\ion{N}{II}$]$ emission within one beam-width of each other.

The agreement between our measurements of $[$\ion{N}{II}$]$ and $[$\ion{C}{I}$]$ with measurements of $[$\ion{H}{I}$]$ suggest a correlation between the Warm Neutral Medium (WNM) associated with $[$\ion{H}{I}$]$ 21\,cm emission and the Warm Ionized Medium (WIM) probed by the $[$\ion{N}{II}$]$ feature (see \citealt{ReviewHaffner}). Correlations between these two phases of the ISM have been shown qualitatively at high altitudes from the galactic plane in spiral galaxies (see \citealt{ReviewHaffner,HartmanBook1997}) and for intermediate to high velocity clouds that are not co-rotating with the disk of the galaxy \citep{Tufte1998,Haffner2001}. Our results suggest that such a correlation exists within the disk of NGC~891. This result is consistent with a clumpy model of the ISM in which neutral condensations exist within a $[$\ion{H}{II}$]$ region with complex interfaces between neutral and ionized regions \citep{ReviewHollenbach,ReviewHaffner}.

\vspace{-18pt}
\section{Conclusion}
The Herschel SPIRE FF includes provision of line frequencies, signal-to-noise ratios, velocity information, and an initial line identification estimate for three HR spectral observations of the edge-on galaxy NGC~891. Using these catalogued measurements we have made rotational measurements of the ionized gas in NGC~891 with $[$\ion{N}{II}$]$ $^3$P$_1$--$^3$P$_0$ fine-structure lines that are detected at high SNR in these observations. We have also included measurements of the narrow neutral carbon energy regime with measurements of the $[$\ion{C}{I}$]$ $^3$P$_1$--$^3$P$_0$ and $[$\ion{C}{I}$]$ $^3$P$_2$--$^3$P$_1$ fine-structure lines.

We present our results in complement to radio measurements of CO and $[$\ion{H}{I}$]$ presented in \citet{Garcia1992,Scoville1993,Sofue1996} and \citet{Kijeong2011} as well as optical measurements of H$\alpha$ presented in \citet{Kamphuis2007}. In spite of the limited spatial and velocity resolution of the SPIRE FTS We have found that the position-velocity profile of $[$\ion{N}{II}$]$ and $[$\ion{C}{I}$]$ lines closely match that of atomic and ionized hydrogen in NGC~891. This result evidences the formation of ionization interfaces on the exterior of neutral clumps in ionization regions within the disk of NGC~891.

This work demonstrates through this simple example that the information collected by the FF and made available through SAFECAT greatly aids the exploitation of observations made with the Herschel SPIRE FTS and enhances the legacy value of the instrument and associated data archive. The wealth of FIR data contained in SAFECAT aids in interpreting a large portion of the highest resolution and most sensitive observations of the FIR universe to date.
\vspace{-30pt}
\section*{Acknowledgements}
\emph{Herschel} is an European Space Agency (ESA) space observatory with science instruments provided by European-led Principal Investigator consortia and with important participation from the National Aeronautics and Space Administration (NASA). This research acknowledges support from ESA, the Canadian Space Agency (CSA), the Canada Research Council (CRC), CMC Microsystems, and the Natural Sciences and Engineering Research Council of Canada (NSERC).

This research has made use of the NASA/IPAC Infrared Science Archive, which is funded by the National Aeronautics and Space Administration and operated by the California Institute of Technology.

This research has made use of the \textsc{SciPy} (\url{www.scipy.org}) and \textsc{Astropy} (\url{www.astropy.org}) Python packages. Table formatting in this paper followed the {\it Planck} Style Guide \citep{PlanckStyle}. 
\nocite{2020SciPy-NMeth, astropy:2013, astropy:2018}\\

\vspace{-22pt}
\section*{Data Availability}
\vspace{-3pt}
The \textit{Herschel} SPIRE Spectral Feature Catalogue has been assigned an ESA Digital Object Identifier (DOI) and is available at: \href{https://doi.org/10.5270/esa-lysf2yi}{doi.org/10.5270/esa-lysf2yi}. The FF code and all FF products are publicly available via the \textit{Herschel} Science Archive.



\vspace{-20pt}
\nocite{chrisThesis}
\bibliographystyle{mnras}
\bibliography{main_Arxivsubmission} 

\appendix
\begin{table*}
\begingroup
\begin{center}
	\caption{\label{tab:fullLineRes} FF results from both sparsely sampled SPIRE FTS observations of NGC 891 (N and S). The variable $\nu$ represents the central frequency of the FF fitted line. For information on line identification and quantum number format see \citetalias{FFlineID}. Transitions that have been manually identified are marked with an asterisk.}
\nointerlineskip
\small
%
\newdimen\digitwidth
\setbox0=\hbox{\rm 0}
\digitwidth=\wd0
\catcode`*=\active
\def*{\kern\digitwidth}
\newdimen\signwidth
\setbox0=\hbox{+}
\signwidth=\wd 0
\catcode`!=\active
\def!{\kern\signwidth}
%
\tabskip=2em plus 2em minus 2em
\halign to \hsize{\hfil#& *#\hfil& *#\hfil& \hfil#\hfil& #\hfill& *#\hfil& *#\hfil& \hfil#\hfil& #\hfill&#\hfil\cr
 &\multispan{8}\hrulefill& \cr
\noalign{\vspace{-8.0pt}}
 &\multispan{8}\hrulefill& \cr
		&\multispan{4}{1342224765(S)} & \multispan{4}{1342224766(N)} \cr
        &Detector & $\nu$ [GHz] & SNR & Transition & Detector & $\nu$ [GHz] & SNR & Transition\cr
        \noalign{\vspace{-5.5pt}}
		&\multispan{8}\hrulefill& \cr
& SLWC3& *\,460.182(61)& 6.0& CO(4--3)& SLWC3& *\,460.349(50)& 14.8& CO(4--3)\cr
& SLWC3& *\,491.196(49)& 6.9& $[$\ion{C}{I}$]$ $^3$P$_1$--$^3$P$_0$& SLWC3& *\,491.470(61)& 7.5& $[$\ion{C}{I}$]$ $^3$P$_1$--$^3$P$_0$ \cr
& SLWC3& *\,574.55(10)& 3.0& CO(5--4)& SLWC3& *\,575.566(63)& 7.2& CO(5--4) \cr
& SLWC3& *\,689.72(11)& 5.3& CO(6--5)& SLWC3& *\,690.521(76)& 10.4& CO(6--5)\cr
& SLWC3& *\,807.43(70)& 13.8& $[$\ion{C}{I}$]$ $^3$P$_2$--$^3$P$_1$& SLWC3& *\,705.80(12)& 5.3& -- \cr
& SLWB3& *\,807.329(70)& 10.1& $[$\ion{C}{I}$]$ $^3$P$_2$--$^3$P$_1$& SLWC3& *\,805.60(22)& 4.3& CO(7--6) \cr
& SLWD1& *\,460.081(62)& 11.2& CO(4--3)$^{\ast}$& SLWC3& *\,808.373(45)& 18.2& $[$\ion{C}{I}$]$ $^3$P$_2$--$^3$P$_1$\cr
& SLWD1& *\,491.096(57)& 11.3& HC$^{18}$O$^+$(6--5)& SLWA3& *\,460.009(81)& 6.8& CO(4--3)\cr
& SLWD1& *\,577.17(19)& 2.8& o-H$_2$CO 8(1,7)--7(1,6)& SLWB3& *\,460.317(72)& 5.6& CO(4--3)\cr
& SLWD1& *\,807.512(50)& 17.1& o-H$_2$CO 12(1,12)--11(1,11)& SLWB3& *\,491.576(45)& 9.2& $[$\ion{C}{I}$]$ $^3$P$_1$--$^3$P$_0$\cr
& SLWD2& *\,460.177(49)& 6.7& CO(4--3)& SLWB3& *\,575.392(74)& 5.5& CO(5--4)\cr
& SLWD2& *\,491.088(51)& 7.2& $[$\ion{C}{I}$]$ $^3$P$_1$--$^3$P$_0$& SLWB3& *\,690.54(11)& 7.2& CO(6--5)\cr
& SLWD2& *\,574.893(60)& 6.1& CO(5--4)& SLWB3& *\,806.26(15)& 6.8& CO(7--6)\cr
& SLWD2& *\,690.380(98)& 7.0& CO(6--5)& SLWB3& *\,808.322(41)& 25.1& $[$\ion{C}{I}$]$ $^3$P$_2$--$^3$P$_1$ \cr
& SLWD2& *\,807.685(53)& 17.6& $[$\ion{C}{I}$]$ $^3$P$_2$--$^3$P$_1$& SLWB4& *\,808.17(0.12)& 6.6& --\cr
& SSWD4& 1\,457.7548(81)& 55.4& \textbf{$[$\ion{N}{II}$]$ $^3$P$_1$--$^3$P$_0$}& SLWC2& *\,460.132(72)& 6.7& CO(4--3)\cr
& SSWD4& 1\,461.989(68) & 6.2& --& SLWD2& *\,808.91(11)& 7.0& -- \cr
& SSWB5& 1\,455.892(71)& 6.7& --& SSWD4& 1\,459.3982(45) &54.9 & \textbf{$[$\ion{N}{II}$]$ $^3$P$_1$--$^3$P$_0$}\cr
& SSWB5& 1\,457.523(22)& 15.0& \textbf{$[$\ion{N}{II}$]$ $^3$P$_1$--$^3$P$_0$}& SSWA4& 1\,459.109(40)& 8.2&\textbf{$[$\ion{N}{II}$]$ $^3$P$_1$--$^3$P$_0$}\cr
& SSWC4& 1\,457.686(10)& 36.6& \textbf{$[$\ion{N}{II}$]$ $^3$P$_1$--$^3$P$_0$}& SSWB5& 1\,458.7231(70)& 60.2& \textbf{$[$\ion{N}{II}$]$ $^3$P$_1$--$^3$P$_0$}\cr
& SSWC5& 1\,457.246(43)& 7.6& \textbf{$[$\ion{N}{II}$]$ $^3$P$_1$--$^3$P$_0$}$^{\ast}$& SSWB5& 1\,465.9(4.4)& -28.1& -- \cr
& SSWD3& 1\,458.010(30)& 12.5& \textbf{$[$\ion{N}{II}$]$ $^3$P$_1$--$^3$P$_0$}& SSWC3& 1\,459.713(34)& 9.5& \textbf{$[$\ion{N}{II}$]$ $^3$P$_1$--$^3$P$_0$}\cr 
& SSWE2& 1\,458.354(14)& 24.5& \textbf{$[$\ion{N}{II}$]$ $^3$P$_1$--$^3$P$_0$}& SSWC4& 1\,459.3070(72)& 50.8& \textbf{$[$\ion{N}{II}$]$ $^3$P$_1$--$^3$P$_0$}\cr
& SSWE3& 1\,457.6126(40)& 73.2& \textbf{$[$\ion{N}{II}$]$ $^3$P$_1$--$^3$P$_0$}& SSWC5& 1\,459.121(14)& 27.7& \textbf{$[$\ion{N}{II}$]$ $^3$P$_1$--$^3$P$_0$}\cr 
& SSWE4& 1\,457.791(44)& 7.9& \textbf{$[$\ion{N}{II}$]$ $^3$P$_1$--$^3$P$_0$}& SSWC5& 1\,557.465(52)& 6.5& -- \cr
& SSWF1& 1\,458.634(13)& 27.8& \textbf{$[$\ion{N}{II}$]$ $^3$P$_1$--$^3$P$_0$}& SSWD3& 1\,459.652(46)& 7.7& \textbf{$[$\ion{N}{II}$]$ $^3$P$_1$--$^3$P$_0$}\cr
& SSWF2& 1\,457.8871(97)& 42.0& \textbf{$[$\ion{N}{II}$]$ $^3$P$_1$--$^3$P$_0$}& SSWE3& 1\,459.3964(71)& 53.1& \textbf{$[$\ion{N}{II}$]$ $^3$P$_1$--$^3$P$_0$}\cr
& SSWG1& 1\,458.120(14)& 24.1& \textbf{$[$\ion{N}{II}$]$ $^3$P$_1$--$^3$P$_0$}$^{\ast}$& SSWF2& 1\,459.532(22)& 16.4& \textbf{$[$\ion{N}{II}$]$ $^3$P$_1$--$^3$P$_0$}\cr
&\multispan{8}\hrulefill& \cr
\noalign{\vspace{-7.5pt}}
}
\end{center}
\endgroup
\vspace{-12pt}
\end{table*}

\begin{table*}
\begingroup
\begin{center}
\newdimen\tblskip \tblskip=5pt
	\caption{\label{tab:fullLineResMap} FF results from the mapping observation of NGC 891 (1342213376, M). Array denotes the hyperspectral cube that the line was detected in and pixel denotes the corresponding pixel row and column of on-sky coordinates of the cube. As in Tab.\,\ref{tab:fullLineRes}, $\nu$ denotes the central line frequency of the fitted line and transitions that have been manually identified are marked with an asterisk.}
\nointerlineskip
\small
%
\newdimen\digitwidth
\setbox0=\hbox{\rm 0}
\digitwidth=\wd0
\catcode`*=\active
\def*{\kern\digitwidth}
\newdimen\signwidth
\setbox0=\hbox{+}
\signwidth=\wd 0
\catcode`!=\active
\def!{\kern\signwidth}
%
\tabskip=2em plus 2em minus 2em
\halign to \hsize{\hfil#& *#\hfil& \hfil#\hfil& *#\hfil& \hfil#\hfil& #\hfill& \hfil#\hfil&\hfil#\hfil& *#\hfil&\hfil#\hfil& #\hfil&#\hfil\cr
 &\multispan{10}\hrulefill& \cr
\noalign{\vspace{-8.0pt}}
 &\multispan{10}\hrulefill& \cr
&	Array & Pixel & $\nu$ [GHz] & SNR & Transition & Array & Pixel & $\nu$ [GHz] & SNR & Transition \cr
\noalign{\vspace{-5.5pt}}
 &\multispan{10}\hrulefill& \cr
 
 & SLW& 0,2& 1\,003.658(78)& 7.7&  --&SSW & 4,7& 1\,457.9463(71)& 57.9& \textbf{$[$\ion{N}{II}$]$ $^3$P$_1$--$^3$P$_0$}\cr
 & SLW& 0,4& *\,807.496(90)& 7.5& $[$\ion{C}{I}$]$ $^3$P$_2$--$^3$P$_1$& SSW & 4,8& 1\,458.146(19)& 20.2& \textbf{$[$\ion{N}{II}$]$ $^3$P$_1$--$^3$P$_0$}\cr
 & SLW& 1,2& *\,578.310(74)& 5.1& --& SSW& 5,4& 1\,458.269(24)& 13.7& \textbf{$[$\ion{N}{II}$]$ $^3$P$_1$--$^3$P$_0$}\cr
 & SLW& 1,3& *\,807.267(97)& 8.1& $[$\ion{C}{I}$]$ $^3$P$_2$--$^3$P$_1$& SSW& 5,5& 1\,458.2241(73)& 44.5& \textbf{$[$\ion{N}{II}$]$ $^3$P$_1$--$^3$P$_0$}\cr
 & SLW& 1,4& *\,575.038(71)& 5.1& CO(5--4)& SSW& 5,6& 1\,458.1745(49)& 56.8& \textbf{$[$\ion{N}{II}$]$ $^3$P$_1$--$^3$P$_0$}\cr
 & SLW& 1,4& *\,807.531(57)& 12.0& $[$\ion{C}{I}$]$ $^3$P$_2$--$^3$P$_1$& SSW& 5,7& 1\,458.259(13)& 32.4& \textbf{$[$\ion{N}{II}$]$ $^3$P$_1$--$^3$P$_0$}\cr
 & SLW& 2,3& *\,491.082(50) & 5.7& $[$\ion{C}{I}$]$ $^3$P$_1$--$^3$P$_0$ & SSW& 5,8& 1\,458.233(42) & 9.0& \textbf{$[$\ion{N}{II}$]$ $^3$P$_1$--$^3$P$_0$}$^{\ast}$ \cr
 & SLW& 2,3& *\,807.772(49)& 13.9& $[$\ion{C}{I}$]$ $^3$P$_2$--$^3$P$_1$& SSW& 6,3& 1\,458.400(61)& 5.4& \textbf{$[$\ion{N}{II}$]$ $^3$P$_1$--$^3$P$_0$}$^{\ast}$ \cr
 & SLW& 2,4& *\,491.064(53)& 5.4& $[$\ion{C}{I}$]$ $^3$P$_1$--$^3$P$_0$& SSW& 6,4& 1\,458.386(16)& 26.8& \textbf{$[$\ion{N}{II}$]$ $^3$P$_1$--$^3$P$_0$}\cr
 & SLW& 2,4& *\,807.763(61)& 12.6& $[$\ion{C}{I}$]$ $^3$P$_2$--$^3$P$_1$& SSW& 6,5& 1\,458.4034(45)& 69.2& \textbf{$[$\ion{N}{II}$]$ $^3$P$_1$--$^3$P$_0$}\cr
 & SLW& 3,2& *\,808.004(91)& 8.7& $[$\ion{C}{I}$]$ $^3$P$_2$--$^3$P$_1$& SSW& 6,6& 1\,458.4324(40)& 81.8& \textbf{$[$\ion{N}{II}$]$ $^3$P$_1$--$^3$P$_0$}\cr
 & SLW& 3,3& *\,460.455(63)& 5.2& CO(4--3)& SSW& 6,7& 1\,458.515(19)& 19.5& \textbf{$[$\ion{N}{II}$]$ $^3$P$_1$--$^3$P$_0$}\cr
 & SLW& 3,3& *\,491.290(49)& 7.4& $[$\ion{C}{I}$]$ $^3$P$_1$--$^3$P$_0$& SSW& 6,8& 1\,458.396(69)& 5.3& \textbf{$[$\ion{N}{II}$]$ $^3$P$_1$--$^3$P$_0$}$^{\ast}$\cr
 & SLW& 3,3& *\,575.227(50)& 6.8& CO(5--4)& SSW& 7,3& 1\,458.605(63)& 5.7& \textbf{$[$\ion{N}{II}$]$ $^3$P$_1$--$^3$P$_0$}$^{\ast}$ \cr
 & SLW& 3,3& *\,690.157(70)& 8.2& CO(6--5)& SSW& 7,4& 1\,458.5916(97)& 41.9& \textbf{$[$\ion{N}{II}$]$ $^3$P$_1$--$^3$P$_0$}\cr
 & SLW& 3,3& *\,805.409(52)& 8.1& CO(7--6)& SSW& 7,5& 1\,458.6415(53)& 59.4& \textbf{$[$\ion{N}{II}$]$ $^3$P$_1$--$^3$P$_0$}\cr
 & SLW& 3,3& *\,807.958(17)& 25.3& $[$\ion{C}{I}$]$ $^3$P$_2$--$^3$P$_1$& SSW& 7,6& 1\,458.6857(80)& 55.1& \textbf{$[$\ion{N}{II}$]$ $^3$P$_1$--$^3$P$_0$}\cr
 & SLW& 3,4& *\,807.994(71)& 12.7& $[$\ion{C}{I}$]$ $^3$P$_2$--$^3$P$_1$& SSW& 7,7& 1\,458.911(32)& 10.3& \textbf{$[$\ion{N}{II}$]$ $^3$P$_1$--$^3$P$_0$}\cr
 & SLW& 4,2& *\,808.249(81)& 9.9& $[$\ion{C}{I}$]$ $^3$P$_2$--$^3$P$_1$& SSW& 8,3& 1\,458.853(32)& 10.1& \textbf{$[$\ion{N}{II}$]$ $^3$P$_1$--$^3$P$_0$}\cr
 & SLW& 4,3& *\,808.152(58)& 14.0& $[$\ion{C}{I}$]$ $^3$P$_2$--$^3$P$_1$& SSW& 8,4& 1\,458.7403(68)& 49.9& \textbf{$[$\ion{N}{II}$]$ $^3$P$_1$--$^3$P$_0$}\cr
 & SLW& 5,2& *\,808.403(70)& 10.5& $[$\ion{C}{I}$]$ $^3$P$_2$--$^3$P$_1$& SSW& 8,5& 1\,458.8363(82)& 43.4& \textbf{$[$\ion{N}{II}$]$ $^3$P$_1$--$^3$P$_0$}\cr
 & SLW& 5,3& *\,808.290(97)& 9.8& $[$\ion{C}{I}$]$ $^3$P$_2$--$^3$P$_1$& SSW& 8,6& 1\,459.012(17)& 24.0& \textbf{$[$\ion{N}{II}$]$ $^3$P$_1$--$^3$P$_0$}\cr
 & SLW& 6,2& *\,808.447(74)& 9.1& $[$\ion{C}{I}$]$ $^3$P$_2$--$^3$P$_1$& SSW& 8,7& 1\,459.134(60)& 5.5& \textbf{$[$\ion{N}{II}$]$ $^3$P$_1$--$^3$P$_0$}$^{\ast}$ \cr
 & SLW& 6,3& *\,808.29(12)& 7.7& --& SSW& 9,3& 1\,458.913(15)& 24.2& \textbf{$[$\ion{N}{II}$]$ $^3$P$_1$--$^3$P$_0$}\cr
 & SSW& 0,7& 1\,457.487(45)& 6.4& \textbf{$[$\ion{N}{II}$]$ $^3$P$_1$--$^3$P$_0$}$^{\ast}$& SSW& 9,4& 1\,458.8414(60)& 53.8& \textbf{$[$\ion{N}{II}$]$ $^3$P$_1$--$^3$P$_0$}\cr
 & SSW& 0,8& 1\,457.501(48)& 6.1& \textbf{$[$\ion{N}{II}$]$ $^3$P$_1$--$^3$P$_0$}$^{\ast}$& SSW& 9,5& 1\,458.9612(87)& 39.0& \textbf{$[$\ion{N}{II}$]$ $^3$P$_1$--$^3$P$_0$}\cr
 & SSW& 1,6& 1\,457.448(37)& 8.7& \textbf{$[$\ion{N}{II}$]$ $^3$P$_1$--$^3$P$_0$}$^{\ast}$& SSW& 9,6& 1\,459.130(53)& 6.3& \textbf{$[$\ion{N}{II}$]$ $^3$P$_1$--$^3$P$_0$}$^{\ast}$\cr
 & SSW& 1,7& 1\,457.522(25)& 12.0& \textbf{$[$\ion{N}{II}$]$ $^3$P$_1$--$^3$P$_0$}& SSW& 10,2& 1\,459.134(50)& 7.4& \textbf{$[$\ion{N}{II}$]$ $^3$P$_1$--$^3$P$_0$}$^{\ast}$ \cr
 & SSW& 1,8& 1\,457.864(35)& 8.7& \textbf{$[$\ion{N}{II}$]$ $^3$P$_1$--$^3$P$_0$}$^{\ast}$& SSW& 10,3& 1\,459.1110(98)& 43.9& \textbf{$[$\ion{N}{II}$]$ $^3$P$_1$--$^3$P$_0$}\cr
 & SSW& 2,6& 1\,457.521(13)& 24.3& \textbf{$[$\ion{N}{II}$]$ $^3$P$_1$--$^3$P$_0$}& SSW& 10,3& 1\,557.240(81)& 6.1& -- \cr
 & SSW& 2,7& 1\,457.6521(94)& 37.1& \textbf{$[$\ion{N}{II}$]$ $^3$P$_1$--$^3$P$_0$}& SSW& 10,4& 1\,458.9989(65)& 47.7& \textbf{$[$\ion{N}{II}$]$ $^3$P$_1$--$^3$P$_0$}\cr
 & SSW& 2,8& 1\,457.977(20)& 17.7& \textbf{$[$\ion{N}{II}$]$ $^3$P$_1$--$^3$P$_0$}& SSW& 10,5& 1\,459.039(18)& 19.3& \textbf{$[$\ion{N}{II}$]$ $^3$P$_1$--$^3$P$_0$}\cr
 & SSW& 2,9& 1\,458.057(56)& 6.4& \textbf{$[$\ion{N}{II}$]$ $^3$P$_1$--$^3$P$_0$}$^{\ast}$& SSW& 11,2& 1\,459.145(10)& 38.4& \textbf{$[$\ion{N}{II}$]$ $^3$P$_1$--$^3$P$_0$}\cr
 & SSW& 3,5& 1\,457.646(35)& 8.9& \textbf{$[$\ion{N}{II}$]$ $^3$P$_1$--$^3$P$_0$}$^{\ast}$& SSW& 11,3& 1\,459.1655(66)& 53.1& \textbf{$[$\ion{N}{II}$]$ $^3$P$_1$--$^3$P$_0$}\cr
 & SSW& 3,6& 1\,457.6681(57)& 56.5& \textbf{$[$\ion{N}{II}$]$ $^3$P$_1$--$^3$P$_0$}& SSW& 11,4& 1\,459.3339(80)& 39.6& \textbf{$[$\ion{N}{II}$]$ $^3$P$_1$--$^3$P$_0$}\cr
 & SSW& 3,7& 1\,457.7984(56)& 58.6& \textbf{$[$\ion{N}{II}$]$ $^3$P$_1$--$^3$P$_0$}& SSW& 11,4& 1\,551.154(83)& 5.5& --\cr
 & SSW& 3,8& 1\,458.043(16)& 18.9& \textbf{$[$\ion{N}{II}$]$ $^3$P$_1$--$^3$P$_0$}& SSW& 11,5& 1\,459.416(34)& 12.3& \textbf{$[$\ion{N}{II}$]$ $^3$P$_1$--$^3$P$_0$}\cr
 & SSW& 4,4& 1\,458.103(65)& 5.2& \textbf{$[$\ion{N}{II}$]$ $^3$P$_1$--$^3$P$_0$}$^{\ast}$& SSW& 12,2& 1\,459.1021(75)& 42.4& \textbf{$[$\ion{N}{II}$]$ $^3$P$_1$--$^3$P$_0$}\cr
 & SSW& 4,5& 1\,457.925(16)& 20.6& \textbf{$[$\ion{N}{II}$]$ $^3$P$_1$--$^3$P$_0$}& SSW& 12,3& 1\,459.1131(72)& 43.1& \textbf{$[$\ion{N}{II}$]$ $^3$P$_1$--$^3$P$_0$}\cr
 & SSW& 4,6& 1\,457.8812(43)& 63.5& \textbf{$[$\ion{N}{II}$]$ $^3$P$_1$--$^3$P$_0$}& SSW& 12,4& 1\,459.349(28)& 13.9& \textbf{$[$\ion{N}{II}$]$ $^3$P$_1$--$^3$P$_0$}\cr
 & & & & & & SSW& 12,5& 1\,459.424(54)& 7.6& \textbf{$[$\ion{N}{II}$]$ $^3$P$_1$--$^3$P$_0$}$^{\ast}$ \cr
&\multispan{10}\hrulefill& \cr
\noalign{\vspace{-7.5pt}}
}
\end{center}
\endgroup
\vspace{-12pt}
\end{table*}

\bsp	
\label{lastpage}
\end{document}